\documentclass[twoside]{article}
\usepackage{qic}
\usepackage{amssymb}

\newcommand{\bra}[1]{\langle {#1} \vert}
\newcommand{\ket}[1]{\vert {#1} \rangle}

\begin{document}
\setlength{\textheight}{8.0truein}    

\runninghead{Linear bosonic quantum channels defined by superpositions $\ldots$}
            {Volkoff}

\normalsize\textlineskip
\thispagestyle{empty}
\setcounter{page}{1}


\vspace*{0.88truein}

\alphfootnote

\fpage{1}

\centerline{\bf
LINEAR BOSONIC QUANTUM CHANNELS DEFINED BY SUPERPOSITIONS}
\vspace*{0.035truein}
\centerline{\bf OF MAXIMALLY DISTINGUISHABLE GAUSSIAN ENVIRONMENTS}
\vspace*{0.37truein}
\centerline{\footnotesize
T.J. VOLKOFF}
\vspace*{0.015truein}
\centerline{\footnotesize\it Department of Physics, Konkuk University}
\baselineskip=10pt
\centerline{\footnotesize\it Seoul 05029, Korea}
\vspace*{0.225truein}

\vspace*{0.21truein}

\abstracts{
A minimal energy quantum superposition of two maximally distinguishable, isoenergetic single mode Gaussian states is used to construct the system-environment representation of a class of linear bosonic quantum channels acting on a single bosonic mode. The quantum channels are further defined by unitary dynamics of the system and environment corresponding to either a passive linear optical element $U_{\mathrm{BS}}$ or two-mode squeezing $U_{\mathrm{TM}}$.  The notion of nonclassicality distance is used to show that the initial environment superposition state becomes maximally nonclassical as the constraint energy is increased. When the system is initially prepared in a coherent state, application of the quantum channel defined by $U_{\mathrm{BS}}$ results in a nonclassical state for all values of the environment energy constraint. We also discuss the following properties of the quantum channels: 1) the maximal noise that a coherent system can tolerate, beyond which the linear bosonic attenuator channel defined by $U_{\mathrm{BS}}$ cannot impart nonclassical correlations to the system, 2) the noise added to a coherent system by the phase-preserving linear amplification channel defined by $U_{\mathrm{TM}}$, and 3) a generic lower bound for the trace norm contraction coefficient on the closed, convex hull of energy-constrained Gaussian states.}{}{}

\vspace*{10pt}


\vspace*{1pt}\textlineskip    
\section{Introduction and Motivation}        

The generation and characterization in the laboratory of quantum superpositions of large amplitude, classical configurations of the electromagnetic field, i.e., quantum superpositions of multimode coherent states, has forcefully demonstrated that the physical principle of quantum superposition can be brought to bear on the kinematics of certain macroscopic systems \cite{harochedavidovich,schoelkopfcat,schoelkopfentangcat,schoelkopfdiss}. Apart from their relevance to foundational studies, macroscopic quantum superpositions of continuous variable (CV) states appear in proposals for parity codes for CV quantum information processing \cite{ralphcats,milburncats}, dissipative generation of quantum error correcting codes \cite{albertcatcodes}, and in the construction of optimal coherent state binary detection protocols \cite{hirota}. On the other hand, in the study of quantum dynamics of CV systems, linear bosonic Gaussian channels serve as testbeds for fundamental concepts, similar to the role of qubit channels in the study of two-level quantum spin systems. At the intersection of the above kinematical and dynamical facets of quantum CV systems lie the problems concerning: 1) linear bosonic Gaussian channels acting on superpositions of maximally distinguishable Gaussian pure states, and 2) linear bosonic (non-Gaussian) quantum channels defined by superpositions of maximally distinguishable Gaussian pure states of the environment. These problems are complementary, and the present paper focuses on the second problem under a physically motivated second moment constraint, corresponding to finite energy (i.e., finite expected photon number).

As we will discuss further in Section \ref{sec:contract}, the distinguishability of two quantum states can be quantified by the distance between the two states with respect to the trace norm. In quantum communication tasks, it is vital to consider sets of quantum signals that are maximally distinguishable with respect to the physical constraints of the communication medium, e.g., the expected energy. In a seminal work, Gottesman, Preskill, and Kitaev \cite{gkp} demonstrated a quantum code that corrects quadrature diffusion errors (e.g., diffusion errors in position or momentum) and which is formed by a countable superposition of orthogonal, infinitely squeezed Gaussian states with equally-spaced mean vectors. In the same work, the authors consider the analogous superpositions of finite energy Gaussian states that are necessary for a physical implementation of the code. Therefore, it appears that superpositions of maximally trace distant, energy constrained Gaussian states are likely to play an important role in error-resistant CV quantum information processing devices. Furthermore, in such devices, the processes of state generation and manipulation will involve linear optics, multimode squeezing, Gaussian measurements, and other dynamical maps acting on several bosonic modes. Therefore, it is of both theoretical and practical importance to understand the properties of quantum channels that are either defined by or take as input superpositions of maximally trace distant, energy-constrained Gaussian states. In the present work, we analyze basic examples of linear bosonic non-Gaussian channels that arise from the coupling of a bosonic mode to an environment mode prepared in a superposition of two maximally trace distant, isoenergetic single mode Gaussian states. Although not necessary for a general analysis, we further restrict the superposition state by requiring that it exhibit minimal energy in a dimension two subspace of the Hilbert space containing the maximally trace distant, isoenergetic single mode Gaussian states.

Recent developments in resource theories of CV quantum coherence \cite{tanvolk,wintergauss,wolfsqueez} provide additional motivation for analyzing the quantum channels that arise in this work. These resource theories quantify the limitations to squeezing, forming superpositions, and entangling CV quantum states under physically motivated classes of dynamics.   


A brief outline of the paper follows. In Section \ref{sec:mathback}, the mathematical setting of linear bosonic non-Gaussian quantum channels is introduced and the $\Xi$ channels are constructed. Section \ref{sec:supdef} contains the definition of the superposition states $\ket{\Omega_{+}}$ of the environment mode. In Section \ref{sec:chananaly}, we analyze the creation of nonclassical states by the $\Xi_{\zeta}$ channel and the noisy dynamics of the $\Xi$ channels. Section \ref{sec:contract} contains a derivation of a lower bound for the contraction coefficient of the $\Xi$ channels on the closed, convex hull of energetically constrained Gaussian states.

\section{\label{sec:mathback}Mathematical background}

Prior work on open quantum dynamics with ``Schr\"{o}dinger cat''-like superpositions of Gaussian states has focused mainly on the quantum properties of a system initially prepared in an even or odd coherent state $\ket{\psi_{\pm}(\alpha)} \propto \ket{-\alpha} \pm \ket{\alpha}$, where $\ket{\alpha}$ is a Glauber coherent state, i.e., an eigenvector of the annihilation operator with eigenvalue $\alpha$ \cite{janszkyevencatbs,dodonovamp,El-Orany2003,filipdegenpar}. However, among state vectors that can be written as a complex linear combination of two Gaussian state vectors $\ket{\psi_{1}}$ and $\ket{\psi_{2}}$ that satisfy $\langle H \rangle_{\ket{\psi}_{1(2)}}=E$, where $H$ is a Hamiltonian quadratic in the creation and annihilation operators, there is nothing particularly special about $\ket{\psi_{\pm}(\alpha)}$. Even the set of states which is considered to be Glauber coherent states, i.e., the union of the kernels of the linear operators $(A-\alpha)$ for $\alpha \in \mathbb{C}$, where $A$, $A^{\dagger}$, $\mathbb{I}$ are generators of the Heisenberg-Weyl Lie algebra $\mathfrak{h}_{3}$ (defined by $[A,\mathbb{I}]=0$, $[A,A^{\dagger}]=\mathbb{I}$), depends on the basis chosen for $\mathfrak{h}_{3}$ \cite{brif}. Physical situations in which the basis of the Heisenberg-Weyl Lie algebra plays an important role include: 1) dynamics of elementary excitations above the Bogoliubov vacuum in the weakly imperfect condensed Bose gas \cite{bogoliubov}, and 2) the Unruh effect \cite{takagi}.

In this work, we define and examine several properties of linear bosonic quantum channels that arise from the linear coupling of a quantum oscillator to a single mode environment, when the initial environment state is prepared in a superposition $\ket{\Omega_{+}}$ of pure, normalized Gaussian states $\ket{\psi_{1}}$, $\ket{\psi_{2}}$ that are maximally distant subject to the energy constraint $\langle a^{\dagger} a \rangle_{\ket{\psi_{1(2)}}} = E$. The relevant pairs of maximally distant, energy-constrained Gaussian states were derived in Ref.\cite{volkoffdist}. Taking $\mathcal{H}_{S(E)}$ to be the Hilbert spaces of the system and environment, both isomorphic to $\ell^{2}(\mathbb{C})$, the quantum channel $\Xi$ is expressed in terms of its Stinespring form
\begin{eqnarray}
\Xi (\rho) &=& \mathrm{tr}_{E}V\rho V^{\dagger} \nonumber \\
V\ket{\psi} &:=&U\ket{\psi}\otimes \ket{\Omega_{+}}_{E} \;  , \; \forall \, \ket{\psi} \in \mathcal{H}_{S},
\label{genchan}
\end{eqnarray}
where $\ket{\Omega_{+}}_{E}\in \mathrm{span}_{\mathbb{C}}\lbrace \ket{\psi_{1}},\ket{\psi_{2}} \rbrace \subset \mathcal{H}_{E}$ is the superposition state of the environment, and $V:\mathcal{H}_{S} \rightarrow \mathcal{H}_{S}\otimes \mathcal{H}_{E}$ is the Stinespring isometry defined by the joint unitary dynamics $U$ of the combined system and environment. Two important examples of linear quantum dynamics correspond to the unitary evolution $U$ defined by: 1) $U_{\mathrm{BS}}(\zeta):=e^{\zeta  a_{0}^{\dagger}a_{1}-\overline{\zeta} a_{0}a_{1}^{\dagger} }$, $\vert \zeta \vert \in [0,\pi/2]$, corresponding to a $SU(2)$ (i.e., passive) linear optical operation implementing an attenuation channel, or 2) $U_{\mathrm{TM}}(r):= e^{r a_{0}a_{1} - r a_{0}^{\dagger}a_{1}^{\dagger}}$, $r>0$, corresponding to an $SU(1,1)$ two-mode squeezing operation implementing a phase-preserving amplification channel. The corresponding channels in Eq.(\ref{genchan}) will be denoted by $\Xi_{\zeta}$ or $\Xi_{r}$, or, when a property applies to both classes of channel, simply by $\Xi$. 

In order to calculate the action of the $\Xi$ channels on quantum states of interest, as will be done in Section \ref{sec:chananaly}, we find it useful to explicitly state the 4$\times$4 symplectic matrices $T_{\mathrm{BS}}(\zeta)$ and $T_{\mathrm{TM}}(r)$ corresponding, respectively, to $U_{\mathrm{BS}}(\zeta)$ and $U_{\mathrm{TM}}(r)$ (via the metaplectic representation \cite{arvind}). Specifically, letting $R:=(q_{0},p_{0},q_{1},p_{1})$ denote the vector of canonical observables, with $q_{j}:= {a_{j}+a_{j}^{\dagger} \over \sqrt{2} }$, $p_{j}:= {-ia_{j}+ia_{j}^{\dagger} \over \sqrt{2} }$, one verifies that $U_{\mathrm{BS}}(\zeta)^{\dagger}RU_{\mathrm{BS}}(\zeta) = RT_{\mathrm{BS}}(\zeta)$ and $U_{\mathrm{TM}}(r)^{\dagger}RU_{\mathrm{TM}}(r) = RT_{\mathrm{TM}}(r)$, where
\begin{equation}
T_{\mathrm{BS}}(\zeta)= 
  \left( {\begin{array}{cccc}
   \cos \zeta & 0 & -\sin \zeta & 0 \\
   0 & \cos \zeta & 0 &  -\sin \zeta\\
   \sin \zeta & 0 & \cos \zeta &  0 \\
   0 & \sin \zeta & 0 &  \cos \zeta
  \end{array} } \right) 
  \label{symplecticmatbs}
\end{equation}
\begin{equation}
T_{\mathrm{TM}}(r)= 
  \left( {\begin{array}{cccc}
   \cosh r & 0 & \sinh r & 0 \\
   0 & \cosh r & 0 &  -\sinh r\\
   \sinh r& 0 & \cosh r &  0 \\
   0 & -\sinh r & 0 &  \cosh r
  \end{array} } \right)   .
\label{symplecticmattm}
\end{equation}
In Eq.(\ref{symplecticmatbs}), we have taken $\zeta \in \mathbb{R}$, which will be the case throughout this work. Then, because any quantum state $\rho$ (of, e.g., mode 0 without loss of generality)  can be expanded over the Weyl CCR $C^{*}$-algebra according to the formula $\rho = {1\over 2\pi}\int \, dxdy\, \chi_{\rho}(x,y)W_{0}(-x,-y)$, where $W_{0}(x,y):=e^{ixq_{0}+iyp_{0}}$ and $\chi_{\rho}(x,y):= \mathrm{tr}\rho W_{0}(x,y)$ is the quantum characteristic function of $\rho$, it follows from Eq.(\ref{genchan}), (\ref{symplecticmatbs}), and (\ref{symplecticmattm}) that
\begin{eqnarray}
\chi_{\Xi_{\zeta}(\rho)}(x,y)&=&\chi_{\rho}( x\cos \zeta , y \cos \zeta ) \chi_{\ket{\Omega_{+}}\bra{\Omega_{+}}}(x\sin \zeta , y \sin \zeta) \nonumber \\
\chi_{\Xi_{r}(\rho)}(x,y)&=&\chi_{\rho}( x\cosh r , y \cosh r  ) \chi_{\ket{\Omega_{+}}\bra{\Omega_{+}}}(x\sinh r  , -y \sinh r).
\label{qucharimage}
\end{eqnarray}

Because $\chi_{\ket{\Omega_{+}}\bra{\Omega_{+}}}: \mathbb{R}^{2}\rightarrow \mathbb{R}$ is not a Gaussian distribution on the plane, the $\Xi$ channels are classified as linear bosonic non-Gaussian channels. The quantum characteristic function of a general class of superposition states that contains $\ket{\Omega_{+}}\bra{\Omega_{+}}$ is computed in Appendix A (see Eq.(\ref{optcharfunct})). Note that if a system is initially prepared in a Gaussian state, the complementary channels corresponding to the $\Xi$ channels are simply special cases of single mode bosonic Gaussian channels \cite{holevoonemode}. Further background on linear bosonic quantum channels, Gaussian states, and relevant concepts from the theory of open bosonic systems can be found in, e.g., Ref.\cite{holevoqubook}.

\section{\label{sec:supdef}Minimal energy superpositions of maximally distant Gaussian environments}

Given two nonorthogonal, normalized pure states $\ket{\psi_{1}}$, $\ket{\psi_{2}}$, contained in the domain of a self-adjoint operator $H$, and that satisfy $\langle H \rangle_{\ket{\psi_{j}}} = E$, we seek a pure state $\ket{\omega_{d,\theta}}\propto \ket{\psi_{1}}+\lambda e^{i\theta}\ket{\psi_{2}}$ defined by parameters $\lambda$, $\theta$, that has minimal expectation value of $H$. We assume that $\langle \psi_{1}\vert \psi_{2}\rangle =:z  \in \mathbb{R}_{+}$ and $\langle \psi_{1}\vert H \vert \psi_{2} \rangle =: c \in \mathbb{R}$. It follows that
\begin{equation}
\langle \omega_{\lambda,\theta} \vert H \vert \omega_{\lambda,\theta} \rangle = {E(1+\lambda^{2})+2\lambda c\cos \theta  \over 1+\lambda^{2}+2z\lambda\cos\theta}.
\end{equation} If $Ez=c$, all quantum states with support contained in $\mathrm{span}_{\mathbb{C}}\lbrace \ket{\psi_{1},}\ket{\psi_{2}} \rbrace$ have the same expectation value for $H$. That this phenomenon arises in nontrivial cases\footnote{A trivial case would be occur if, e.g., $\ket{\psi_{1}}$ and $\ket{\psi_{2}}$ were degenerate, orthogonal eigenvectors of $H$.}  can be verified by taking $H=a^{\dagger}a$, $E \in \mathbb{N}$, and $\ket{\psi_{1}}=\ket{\alpha}$ to be a Heisenberg-Weyl coherent state with $\alpha = E$, and $\ket{\psi_{2}} = \ket{n}$ to be a Fock state with $n=E$. For $Ez\neq c$, the minimal value of $\langle H \rangle_{\ket{\omega_{\lambda,\theta}}}$ is obtained on the subset of states $\ket{\omega_{\lambda=1,\theta}}$. In particular, if $Ez > c$ ($Ez<c$), then $\ket{\omega_{\lambda=1,\theta=0}}$ ($\ket{\omega_{\lambda=1,\theta=\pi}}$) attains the minimal value of $\langle H \rangle_{\ket{\omega_{\lambda,\theta}}}$.   

To explore the properties of a non-Gaussian quantum environment prepared in a superposition of maximally distant isoenergetic Gaussian pure states, it is first necessary to find a pair of Gaussian pure states $\ket{\varphi_{1}}$, $\ket{\varphi_{2}}$ that exhibit minimal fidelity subject to the energy constraint $\langle \varphi_{j} \vert a^{\dagger}a \vert \varphi_{j} \rangle = E$, $j=1,2$ \cite{volkoffdist}. Using the definitions $S(z):= e^{{1\over 2}\left( \overline{z} a^{2} - za^{\dagger 2} \right)}$ for the unitary squeezing operator and $D(\alpha):=e^{\alpha a^{\dagger} - \overline{\alpha}a }$ for the unitary displacement operator, and defining $\ket{(\alpha,z)} := D(\alpha)S(z)\ket{0}$, where $\ket{0}$ is the Fock vacuum, a  pair of minimal fidelity, isoenergetic Gaussian states (unique up to U(1) phase shifts) is given by $\ket{ \left( \pm r(d_{c}(E)),{1\over 2}\ln d_{c}(E) \right)}$, where
\begin{eqnarray}
d_{c}(E)&=&2E+1 \nonumber \\
r(d_{c}(E))&=&\sqrt{E-{1\over 4}\left( {d_{c}(E)-1\over \sqrt{d_{c}(E)}}\right)^{2}}= \sqrt{{E^{2}+E\over 2E+1}}.
\end{eqnarray} Furthermore, it follows from the adjoint action of the squeezing operator and displacement operator on the creation and annihilation operators that the states $\ket{ \left( \pm r(d_{c}(E)),{1\over 2}\ln d_{c}(E) \right)}$ satisfy $Ez>c$. Therefore, the state $\ket{\Omega_{+}} \propto \ket{ \left(  r(d_{c}(E)),{1\over 2}\ln d_{c}(E) \right)} + \ket{ \left( - r(d_{c}(E)),{1\over 2}\ln d_{c}(E) \right)}$ is the lowest energy pure state in the linear span of $\ket{ \left( \pm r(d_{c}(E)),{1\over 2}\ln d_{c}(E) \right)}$. The relation $D(\alpha)S(z)=S(z)D(\alpha e^{z})$ for real $\alpha$ and $z$ implies that $\ket{\Omega_{+}}$ is a squeezed version of the even coherent state \cite{dodonov}. Explicitly, \begin{equation} \ket{\Omega_{+}}\propto S({1\over 2}\ln d_{c}(E))\left( \ket{\gamma_{c}} \pm \ket{-\gamma_{c}} \right) , \label{stateform}\end{equation} where $\gamma_{c}:=r(d_{c}(E))\sqrt{d_{c}(E)}$. In the following sections, we take $\ket{\Omega_{+}}$ to be the initial state of the environment that defines the channel $\Xi$ in Eq.(\ref{genchan}), and make use of both expressions in Eq.(\ref{stateform}), according to convenience in the application. Regardless of whether Stinespring form of the $\Xi$ channel is defined by $U_{\mathrm{BS}}$ or $U_{\mathrm{TM}}$, it exhibits the $\mathbb{Z}_{2}$ covariance property $e^{i\pi a^{\dagger}a}\Xi (\rho)e^{-i\pi a^{\dagger}a} = \Xi ( e^{i\pi a^{\dagger}a}\rho e^{-i\pi a^{\dagger}a})$ that is inherited from the $\mathbb{Z}_{2}$ symmetry of the environment state $\ket{\Omega_{+}}$.  Further technical facts concerning the state $\ket{\Omega_{+}}$ and the channel $\Xi$, which will be useful in subsequent sections, are collected in Sections \ref{sec:app1} and \ref{sec:app2} (see also Fig.\ref{fig:omegaprops}).

\section{Creation and destruction of nonclassicality via the $\Xi$ channel\label{sec:chananaly}}

The 50:50 beamsplitter that defines the Stinespring dilation of the $\Xi_{\zeta=\pi/4}$ channel is capable of generating entanglement between two optical modes. Possibly the most celebrated example of this phenomenon is the generation of 1 ebit of entanglement in the Hong-Ou-Mandel effect \cite{PhysRevLett.59.2044}, which can be succinctly stated as \begin{equation}
U_{\mathrm{BS}}(\pi/4)\ket{1}\otimes \ket{1} = {1\over \sqrt{2}}\left( \ket{0}\otimes \ket{2} + \ket{2} \otimes \ket{0} \right)
\end{equation}
The entanglement generated by the Stinespring isometry of a quantum channel is responsible for the noise that the channel imparts to an input quantum state. In general, it is known that nonclassicality of the input state is necessary for entanglement generation by a 50:50 beamsplitter \cite{PhysRevA.65.032323,PhysRevA.88.044301}. Before analyzing the noisy quantum dynamics of the $\Xi$ channel in Sections \ref{sec:ndist}, \ref{sec:noisy}, and \ref{sec:noiseamp}, we consider here the correlations that are created between a vacuum port and a $\ket{\Omega_{+}}$ port by a 50:50 beamsplitter $U_{BS}(\pi/4)$. In particular, to quantify the correlation between the modes that gives rise to the noise generated by the $\Xi_{\zeta = \pi/4}$ channel, we use the R\'{e}nyi $\alpha = 2$ entropy defined as  \cite{PhysRevLett.104.157201,PhysRevLett.109.190502} \begin{equation} S_{\alpha = 2}(\rho):= -\log_{2}\mathrm{tr}_{S}\left( (\mathrm{tr}_{E}\rho)^{2} \right)\end{equation} for $\rho$ a quantum state on $\mathcal{H}_{S}\otimes \mathcal{H}_{E}$. $S_{\alpha=2}(\rho)$ is a lower bound on the entanglement entropy $S_{\alpha = 1}(\rho):= -\mathrm{tr}_{S}\left( \mathrm{tr}_{E}\rho \log_{2} \mathrm{tr}_{E}\rho \right)$.

\begin{figure} [t!]
\centerline{\epsfig{file=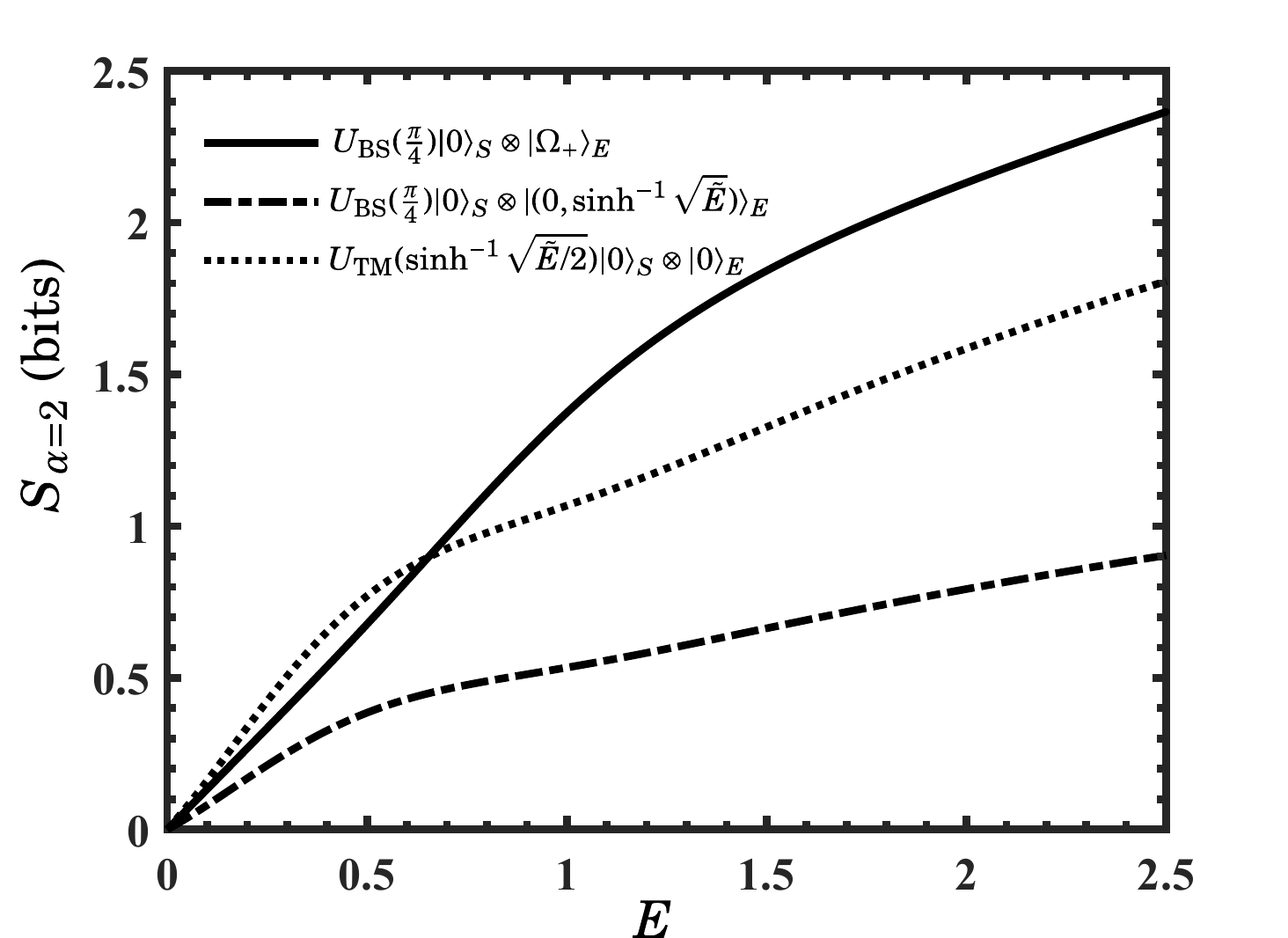, width=8.2cm}} 
\vspace*{13pt}
\fcaption{\label{fig:renyi2} R\'{e}nyi $\alpha =2$ entanglement entropy for three states with respect to the energy constraint $E$ that defines $\ket{\Omega_{+}}$. We define $\tilde{E}:= \langle a^{\dagger}a \rangle_{\ket{\Omega_{+}}}$. For a given value of $E$, all states have the same expected energy $\tilde{E}$.}
\end{figure}

In Fig.\ref{fig:renyi2}, $S_{\alpha = 2}$ is plotted for three states of $\mathcal{H}_{S}\otimes \mathcal{H}_{E}$ with the same expected energy $\tilde{E}:= \langle a^{\dagger}a \rangle_{\ket{\Omega_{+}}}$ incident on a 50:50 beamsplitter: 1) $\ket{0}_{S}\otimes \ket{\Omega_{+}}_{E}$, 2)  $\ket{0}_{S} \otimes S(\sinh^{-1}\sqrt{\tilde{E}})\ket{0}_{E}$, and 3) $S(\sinh^{-1}\sqrt{\tilde{E}/2})\ket{0}_{S} \otimes S(-\sinh^{-1}\sqrt{\tilde{E}/2})\ket{0}_{E}$. In the first two cases, computation of beamsplitter dynamics was greatly simplified by use of the following state-valued Gaussian integral expression for the squeezed state $S(w)\ket{0}$ \cite{janszky} (where we take $w \in \mathbb{R}_{+}$ for simplicity):
\begin{equation}
S(w)\ket{0} = {e^{w/2}\over \sqrt{\pi (e^{2w}-1)}}\int_{-\infty}^{\infty}dx \, e^{-{1\over e^{2w}-1}x^{2}}D(ix)\ket{0},
\end{equation} and the analogous expression for $\ket{\Omega_{+}}$ (see Appendix \ref{sec:altexpr}). In the third case, we note that $U_{BS}(\pi/4)S(\sinh^{-1}\sqrt{\tilde{E}/2})\ket{0}_{S} \otimes S(-\sinh^{-1}\sqrt{\tilde{E}/2})\ket{0}_{E} = U_{TM}(\sinh^{-1}\sqrt{\tilde{E}/2})\ket{0}_{S}\otimes \ket{0}_{E}$.

For all energies, we reach the conclusion that a quantum environment mode prepared in the state $\ket{\Omega_{+}}$ allows for creation of greater mode entanglement by a linear optical coupling to a classical mode than can be achieved with a squeezed vacuum environment mode of the same energy. Furthermore, for $E\gtrsim 0.7$, the system-environment correlations of $U_{BS}(\pi/4)\ket{0}_{S}\otimes\ket{\Omega_{+}}_{E}$ are greater than those of a two-mode squeezed state of the same energy, which is a prototypical example of entangled Gaussian state.

\subsection{\label{sec:ndist}Nonclassicality distance induced by $\Xi_{\zeta=\pi/4}$ attenuator channel}
A quantum oscillator, initially prepared in a classical state\footnote{As is customary in the quantum optics literature, we consider a \textit{classical state} $\rho$ of a quantum oscillator to be defined by a Borel measure $\mu_{\rho}(d^{2}\alpha)$ on $\mathbb{C}$ via the relation $\rho = \int \mu_{\rho}(d^{2}\alpha) \vert \alpha \rangle \langle \alpha \vert $ \cite{hillerypureclass}.}$\,$  and subsequently coupled by 50:50 beamsplitter to a single-mode environment prepared in a superposition of macroscopically distinct Gaussian pure states, does not necessarily evolve to a state that exhibits substantial nonclassical character.
This can be shown by quantifying the nonclassicality of a state $\rho$ by the minimal distance from the set of classical states, i.e., by $\delta(\rho) := \inf_{\sigma \in \Delta}\Vert \rho - \sigma \Vert_{1}$, where $\Vert \cdot \Vert_{1}$ is the trace norm and $\Delta$ is the set of classical states \cite{PhysRevA.35.725}. In the simple scenario in which the quantum oscillator is taken to be Fock vacuum (which trivially satisfies $\delta(\ket{0}\bra{0}) =0$) and the environment mode is initialized in the even coherent state $\ket{\psi_{+}(\alpha)} \propto \ket{\alpha} + \ket{-\alpha}$ (which satisfies $\delta(\ket{\psi_{+}(\alpha)}\bra{\psi_{+}(\alpha)}) \le 2e^{-\alpha^{2}}\sinh(\alpha^{2}) < 1$ \cite{nair}), application of a 50:50 beamsplitter produces an entangled coherent state $\propto \ket{{\alpha\over \sqrt{2}}}_{S}\otimes \ket{{\alpha\over \sqrt{2}}}_{E} + \ket{-{\alpha\over \sqrt{2}}}_{S}\otimes \ket{-{\alpha\over \sqrt{2}}}_{E}$. Taking the partial trace over the environment results in a system state $\rho '$ that satisfies:
\begin{eqnarray}
\delta(\rho ') &\le &\Vert \rho ' - {1\over 2}\sum_{j=0}^{1}\big\vert {(-1)^{j}\alpha \over \sqrt{2}} \big\rangle \big\langle  {(-1)^{j}\alpha \over \sqrt{2}} \big\vert \Vert_{1} \nonumber \\ &= & e^{-\alpha^{2}}\tanh \alpha^{2} ,
\end{eqnarray}
with the right hand side attaining a maximum of $\sim 0.3003$ at $\alpha = \sqrt{ {1\over 2}\sinh^{-1}2} $. For  $\alpha \rightarrow \infty$ ($\alpha \rightarrow 0$) the final system state $\rho '$ is exponentially (polynomially) close to classical. Replacing the 50:50 beamsplitter in this example by a linear amplifier results in a classical output state of the system for any value of the gain.

If the environment is prepared in a squeezed vacuum state and coupled via 50:50 beamsplitter to a system prepared in Fock vacuum, the reduced state of the system is classical for all values of the squeezing parameter of the environment mode. Note that as the squeezing parameter of a squeezed state goes to infinity, its trace distance from the set of classical states increases toward 2. Therefore, although a squeezed state is asymptotically orthogonal to the set of classical states, a squeezed environment cannot be used to increase the nonclassicality of a system by linear optical coupling.

\begin{figure} [t!]
\centerline{\epsfig{file=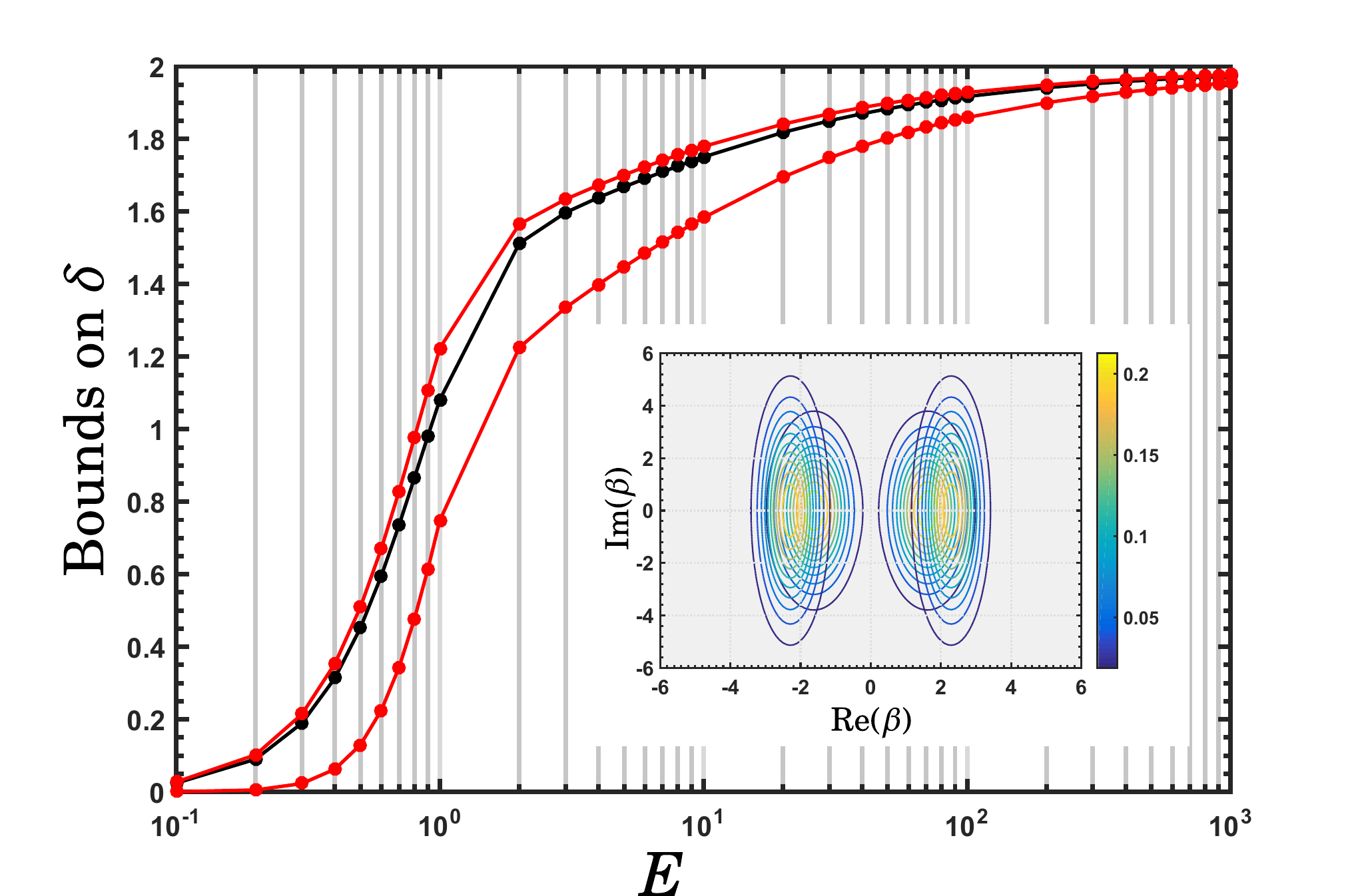, width=8.2cm}} 
\vspace*{13pt}
\fcaption{\label{fig:nonclass} Semilog plot of the bounds on $\delta(\rho)$ in Eq.(\ref{bschanbound}) for $\rho = \ket{\Omega_{+}}\bra{\Omega_{+}}$ (red circles), and the upper bound for $\delta(\rho)$ when $\rho = \Xi_{\zeta = \pi/4}(\ket{0}\bra{0})$ (black circles). The inset shows contour plots of $\vert \langle \beta \vert \Omega_{+} \rangle \vert^{2}$ (ellipses centered farther from 0) and $\langle \beta \vert \Xi(\ket{0}\bra{0} ) \vert \beta \rangle$ (ellipses centered closer to 0) for $E=10$.}
\end{figure}

By contrast, the state $\ket{\Omega_{+}}$, the value $\delta ( \ket{\Omega_{+}}\bra{\Omega_{+}} ) $ approaches the maximal possible value 2 as the energy $E$ becomes infinite and can define an attenuator channel that creates a nonclassical system state from a classical system state. Bounds on $\delta ( \ket{\Omega_{+}}\bra{\Omega_{+}} ) $ can be calculated from the following inequalities \cite{nair}:
\begin{eqnarray}
\delta \left( \ket{\Omega_{+}}\bra{\Omega_{+}} \right) &\ge& 2\left( 1- \sup_{\beta \in \mathbb{C}} \vert \langle \beta \vert \Omega_{+} \rangle \vert^{2}  \right) \nonumber \\ \delta \left( \ket{\Omega_{+}}\bra{\Omega_{+}} \right) &\le& 2\sqrt{ \left( 1- \sup_{\beta \in \mathbb{C}} \vert \langle \beta \vert \Omega_{+} \rangle \vert^{2} \right)  }   .
\label{bschanbound}
\end{eqnarray}
These bounds are shown in Fig.\ref{fig:nonclass} (red lines). It can be seen that $\delta (\ket{\Omega_{+}}\bra{\Omega_{+}})$ grows at least like a power of $E$ for $E\lesssim 1$.

The distance $\delta ( \Xi_{\zeta}(\ket{0}\bra{0}))$  possesses an upper bound given by \begin{equation}
\delta( \Xi_{\zeta}(\ket{0}\bra{0})\le 2\left( 1-\sup_{\beta \in \mathbb{C}}\langle \beta \vert \Xi_{\zeta}(\ket{0}\bra{0} ) \vert \beta \rangle \right)^{1/2},
\end{equation}  which is shown in Fig.\ref{fig:nonclass} for $\zeta = \pi/4$ (black circles). However, for $\xi = \pi/4$, a tighter upper bound is given by $ \delta ( \Xi_{\zeta=\pi/4}(\ket{0}\bra{0}) )\le \delta (\ket{\Omega_{+}}\bra{\Omega_{+}} )$. This inequality follows from the fact that quantum channels that map classical states to classical states decrease the distance $\delta$ \cite{nair}. Although $\Xi_{\zeta}$ does not preserve the set of classical states, the state $\Xi_{\zeta = \pi/4}(\ket{0}\bra{0})$ has an alternative expression as the image of $\ket{\Omega_{+}}\bra{\Omega_{+}}$ under a bosonic Gaussian channel that maps classical states to classical states; explicitly, $\Xi_{\zeta = \pi/4}(\ket{0}\bra{0}) = \mathrm{tr}_{E} U_{\mathrm{BS}}(\zeta = \pi/4)\ket{\Omega_{+}}\bra{\Omega_{+}} \otimes \ket{0}_{E}\bra{0}_{E} U_{\mathrm{BS}}(\zeta = \pi/4)^{\dagger}$.

A general lower bound for $\delta(\rho)$ when $\rho$ satisfies the inequality $\mathrm{sup}_{\beta \in \mathbb{C}}\langle \beta \vert \rho \vert \beta \rangle \le \mathrm{tr}(\rho^{2})$ was derived in Ref.\cite{PhysRevA.35.725}, but is not applicable to the state $\Xi_{\zeta = {\pi \over 4}}(\ket{0}\bra{0} )$. To the knowledge of the author, a general lower bound for $\delta (\rho)$ when $\rho$ violates that inequality is lacking. However, the following Proposition guarantees that $\Xi_{\zeta = {\pi \over 4}}(\ket{0}\bra{0} )$ is nonclassical in the limit of large energy constraint $E$. 
The proof makes use of the fact that a quantum state $\rho$ is classical if and only if $\vert \chi_{\rho}(x,y) \vert \le e^{-{1\over 4}(x^{2}+y^{2})}$ everywhere on the plane \cite{agarwalbook}. 

\textbf{Proposition 2.} \textit{For all coherent states} $\ket{\alpha}$, $\alpha \in \mathbb{C}$\textit{, the following inequality holds:}\begin{equation}\lim_{E \rightarrow \infty} \delta(\Xi_{\zeta= {\pi \over 4}}(\ket{\alpha}\bra{\alpha}) )>0 . 
\end{equation}

\textbf{Proof of Proposition 2.} We compare the $E\rightarrow \infty$ behavior of the characteristic function $g(x,y):=\chi_{\Xi_{\zeta= {\pi \over 4}}(\ket{\alpha}\bra{\alpha})}(x,y)$, which can be computed using Eq.(\ref{qucharimage}) and Eq.(\ref{optcharfunct}), to the Gaussian $v(x,y):= e^{-{1\over 4}(x^{2}+y^{2})}$. Because $\vert g(x,y) \vert$ does not depend on the mean vector of $\ket{\alpha}$, we can specialize to $\alpha = 0$ (i.e., take vacuum input). As $E\rightarrow \infty$, $g(x,y)$ is asymptotically equal to $f(x,y)$, where
\begin{equation}
f(x,y):= e^{-{1\over 8}(x^{2}+y^{2})}e^{-{1\over 8}(e^{-2w}x^{2}+e^{2w}y^{2})}\cos \left( {e^{w}x \over 2} \right) ,
\end{equation}
and where $w:={1\over 2}\ln d_{c}(E) > 0$. This asymptotic behavior of $g(x,y)$ follows from the fact that $\lim_{E \rightarrow \infty}{\gamma_{c} \over d_{c}(E)} = {1\over 2}$. Let us now assume that as $E \rightarrow \infty$ (i.e., as $w \rightarrow \infty$), $ \vert g(x,y) \vert \le v(x,y)$  for all $(x,y)\in \mathbb{R}^{2}$. Then, on the $y=0$ line, it follows that 
$\vert e^{-{e^{-2w}\over 8} x^{2} }\cos \left( {e^{w}x \over 2} \right) \vert \le e^{-{1\over 8}x^{2}}$  for all $x$ and, therefore,
\begin{equation}
\big\vert \cos \left( {e^{w}x \over 2} \right) \big\vert  \le e^{-{1\over 8}(1-e^{-2w})x^{2}}
\label{intermed1}
\end{equation}
for all $x$. But, for any $w > 0$, Eq.(\ref{intermed1}) does not hold for all $x$ and we conclude that $\lim_{E \rightarrow \infty} \delta(\Xi_{\zeta= {\pi \over 4}}(\ket{\alpha}\bra{\alpha}) ) \neq 0$. $\square$

Therefore, whereas an asymptotically classical reduced system state is obtained when a 50:50 beamsplitter couples the system Fock vacuum to an environment port prepared in an even coherent state or in a squeezed vacuum state, an asymptotically nonclassical reduced system state is obtained when the environment is prepared in the state $\ket{\Omega_{+}}$.

\subsection{\label{sec:noisy}Nonclassicality imparted by a noisy $\Xi_{\zeta}$ attenuator channel}

We now proceed to address the following two questions regarding the potential of the channel $\Xi_{\zeta}$ to create nonclassical states in the presence of noise: 1) What is the thermal noise threshold that a quantum oscillator, initially prepared in a coherent state, can tolerate, above which the channel $\Xi_{\zeta}$ outputs a classical state? and, 2) What is the minimal amount of thermal noise $N$ that must be added to the environment state $\ket{\Omega_{+}}$ to guarantee that the channel $\rho \mapsto \mathrm{tr}_{E}\left( U_{\mathrm{BS}}(\zeta)\rho \otimes \Phi_{N}(\ket{\Omega_{+}}_{E}\bra{\Omega_{+}}_{E} ) U_{\mathrm{BS}}(\zeta)^{\dagger} \right)$ maps all classical states into classical states, where $\Phi_{N}$ is a classical Gaussian noise channel (type $B_{2}$ of Ref.\cite{holevoonemode})? When $\zeta = \pi/4$, i.e., $U_{\mathrm{BS}}(\zeta)$ is the 50:50 beamsplitter, the critical noise values that answer these questions are  equal and we therefore address only the first question.

The effect of thermal noise on the system can be modeled by the action of the bosonic Gaussian channel $\Phi_{N}$ that imparts classical noise $N$ to a coherent state $\ket{\beta}$ of the system via $\Phi_{N}(\ket{\beta}\bra{\beta}) = \int {d^{2}z \over \pi N}e^{-{\vert z - \beta \vert^{2} \over N}}\ket{z}\bra{z}$. Without loss of generality in the consideration of coherent states of the system, we restrict to the case of a system initially prepared in the Fock vacuum $\beta = 0$. Then, with $q:= (a+a^{\dagger}) / \sqrt{2}$, $p:= (-i a+ i a^{\dagger}) / \sqrt{2}$ the position and momentum quadratures, respectively, the characteristic function $\chi_{\rho '}(x,y):= \mathrm{tr} \left( \rho ' e^{ixq + iyp} \right)$ of the output state of the channel, $\rho ':= \Xi_{\zeta} \circ \Phi_{N}(\ket{0}\bra{0})$, is given by (see Eq.(\ref{qucharimage})) 

\begin{eqnarray}
\chi_{\Xi_{\zeta} \circ \Phi_{N}(\ket{0}\bra{0})}(x,y) &=& e^{-{\cos^{2} \vert \zeta \vert \over 2}(N+{1\over 2}) (x^{2} + y^{2})} \nonumber \\ 
&{}& \cdot \chi_{\ket{\Omega_{+}}\bra{\Omega_{+}}}( x\sin \vert \zeta \vert,y \sin \vert \zeta \vert)
\end{eqnarray}
where $\chi_{\ket{\Omega_{+}}\bra{\Omega_{+}}}$ is given by Eq.(\ref{optcharfunct}). Hereafter, we take $ \zeta  = \pi/4$, which corresponds to the 50:50 beamsplitter.

We again recall that a state $\rho$ is classical if and only if $\vert \chi_{\rho}(x,y) \vert \le e^{-{1\over 4}(x^{2}+y^{2})}$ everywhere on the plane \cite{agarwalbook}. It follows that for $\rho '$ to be classical, it is necessary that there does not exist $x \in \mathbb{R}$ such that
\begin{equation}
e^{-{1\over 4}x^{2}} - \big\vert \chi_{\rho '}(x,0) \big\vert <0 .
\label{nonclassconst}
\end{equation}
By using Eq.(\ref{optcharfunct}) and  Eq.(\ref{stateform}), and defining $c(x):= \vert \cos \left( {\gamma_{c} x \over \sqrt{d_{c}(E)}} \right) + e^{-2\gamma_{c}^{2}} \vert / (1+ e^{-2\gamma_{c}^{2}})$,  it follows  that
\begin{equation}
e^{-{1\over 4}x^{2}} - \big\vert \chi_{\rho '}(x,0) \big\vert =  e^{-{1\over 4}x^{2}} -e^{ -{(N+{1\over 2})\over 4}x^{2}}e^{-{x^{2}\over 8d_{c}(E)}}c(x).
\label{diffpre}
\end{equation}
Note that $c(x)\le 1$ for all $x$. For any $\epsilon > 0$, taking the classical noise value $N = {1\over 2}-{1 \over 2d_{c}(E)} - \epsilon $ results in the simplification of the right hand side of Eq.(\ref{diffpre}) to the expression $e^{-{1\over 4}x^{2}}\left( 1-e^{{\epsilon x^{2} \over 4}}c(x) \right)$ which cannot be greater than zero for all $x$.
Therefore, the state $\Xi_{\zeta = \pi/4}\circ \Phi_{N}(\ket{\beta}\bra{\beta})$ is nonclassical for $N < N_{\mathrm{crit}}$, where
\begin{equation}
N_{\mathrm{crit}} = {1\over 2} - {1\over 2 d_{c}(E)}.
\label{crit}
\end{equation}
For all environment energy constraints $E$, it also holds that for $N \ge N_{\mathrm{crit}}$, $\Xi_{\zeta = \pi/4}\circ \Phi_{N}(\ket{\beta}\bra{\beta})$ is classical. This is readily confirmed analytically in the $E\rightarrow 0$ and $E\rightarrow \infty$ limits and can be confirmed numerically for intermediate values.

The critical threshold of thermal noise $N$ that a system prepared in a state $\rho$ can tolerate, above which the state $\Phi_{N}(\rho)$ is classical, has been analyzed under the name ``nonclassicality depth'' of the state $\rho$ \cite{leenonclassdepth}. This notion has been further extended to characterize the minimal classicalizing noise for arbitrary sets of states of multimode bosonic systems, and to characterize the nonclassicality breaking of general quantum channels \cite{sabapathynonclass}.

\subsection{\label{sec:noiseamp}Output noise of the $\Xi_{r}$ amplifier channel}

Any quantum channel carrying out phase-preserving linear amplification of a single-mode bosonic signal can be expressed in the Stinespring form with the unitary dynamics of system and environment given by the two-mode squeeze operator $U=U_{TM}(r)$ \cite{cavesamp}. The ideal phase-preserving linear amplifier which, by definition, exhibits minimal noise in the second moment of the electric field of the output state, can be achieved by preparation of the environment mode in the Fock vacuum state. For the $\Xi_{r}$ channel, we are interested in the second moment noise generated when the environment mode is prepared in $\ket{\Omega_{+}}$.


We quantify the mean second moment noise of a state $\rho$ by $\nu(\rho):= {1\over \pi}\int_{0}^{\pi}d\theta \langle \left( \Delta x^{(\theta)} \right)^{2} \rangle_{\rho}$, where $x^{(\theta)}:= e^{i\theta a^{\dagger} a}qe^{-i\theta a^{\dagger} a}$. Note that for an energy constraint $\langle a^{\dagger}a \rangle_{\rho} = \epsilon$, the mean second moment noise is maximized in the set of states that satisfy $\langle a \rangle_{\rho}=0$. In order to compare the mean second moment noise of an input state $\rho$ to the mean second moment noise of the output $\Xi_{r}(\rho)$, we form the quantity $\mu(\rho) := {1\over g^{2}}(\nu (\Xi_{r} (\rho)) / \nu (\rho)) \ge {1\over g^{2}}$, in which a factor of the squared amplifier gain $g^{2}:= \cosh^{2}r$ is canceled in the noise ratio. For an input coherent state, one finds that $\nu(\ket{\beta}\bra{\beta})=2\left( 1-{1\over g^{2}}\right)\left( 1+ \langle a^{\dagger}a \rangle \right) + \mathcal{O}({1\over g^{2}})$, whereas for the input state $\ket{\Omega_{+}}$, $\nu(\ket{\Omega_{+}}\bra{\Omega_{+}})={\left( 2-{1\over g^{2}}\right)\langle a^{\dagger}a \rangle_{\ket{\Omega_{+}}} + 1 + \mathcal{O}({1\over g^{2}}) \over \langle a^{\dagger}a \rangle_{\ket{\Omega_{+}}} + 1} $. The $\Xi_{r}$ channel is able to impart extensive phase insensitive second moment noise to an input coherent state, but not to a $\ket{\Omega_{+}}$ input, which already exhibits extensive mean second moment noise. We note that the channel $\Xi_{r}$ maps the set of classical states to classical states, so the analysis of Section \ref{sec:noisy} is not relevant here.


\section{Contraction of a diameter by the $\Xi$ channel\label{sec:contract}}

In Section \ref{sec:supdef}, a trace norm diameter of the set $CCH(G(\mathcal{H})_{E})$, where $CCH(G(\mathcal{H})_{E})$ denotes the closed, convex hull of the set of Gaussian states $\sigma$ that satisfy $\langle a^{\dagger}a \rangle_{\sigma} = E$, was used to construct $\ket{\Omega_{+}}$ and, subsequently, the $\Xi$ channels. Because the trace distance has an operational characterization in terms of the minimal probability of error in the task of distinguishing two quantum states which have equal \textit{a priori} probabilities \cite{barnettcroke,helstrombook,bae}, and because a maximum of the continuous, jointly convex function $(\rho , \sigma)\mapsto \Vert \rho - \sigma \Vert_{1}$ is necessarily attained on a pair of extreme points (viz., pure states), it follows that the state $\ket{\Omega_{+}}$ is a superposition of maximally distinguishable states in $CCH(G(\mathcal{H})_{E})$.

Furthermore, the trace distance data processing inequality \cite{wildebook} states that for two quantum states $\rho$ and $\sigma$, and a completely positive, trace preserving map $\Phi$, \begin{equation}\Vert \Phi(\rho - \sigma) \Vert_{1}\le \Vert \rho - \sigma \Vert_{1}.\end{equation} Therefore, the minimal loss of binary distinguishability of two quantum signals in the set $CCH(G(\mathcal{H})_{E})$ when transmitted through the $\Xi$ channel with equal \textit{a priori} probability is characterized by the contraction coefficient
\begin{equation}
\tau(\Xi):= \mathrm{max}_{\rho , \sigma \in CCH(G(\mathcal{H})_{E})   }{\Vert \Xi(\rho)-\Xi(\sigma)\Vert_{1} \over \Vert \rho - \sigma\Vert_{1}}.
\label{tau}
\end{equation}
If $\tau (\Xi)$ is close to its maximal value 1, then there exist two quantum signals in $CCH(G(\mathcal{H})_{E})$ that can be passed through the channel $\Xi$ without a large loss in the maximal probability of successfully distinguishing the signals (over all possible measurements). In the absence of the restriction to $CCH(G(\mathcal{H}))$, the trace norm contraction coefficient is achieved on a pair of orthogonal pure states \cite{ruskaitr}. In the present case, we calculate a lower bound to $\tau(\Xi)$ which makes use of its $\mathbb{Z}_{2}$ covariance property. Consider the states $\rho_{1}$, $\rho_{2} \in CCH(G(\mathcal{H}))$ such that $e^{i\pi a^{\dagger}a}\rho_{1}e^{-i\pi a^{\dagger}a} = \rho_{2}$. From the definition in Eq.(\ref{tau}) and $\mathbb{Z}_{2}$ covariance, it follows that
\begin{eqnarray}
\tau(\Xi) &\ge & {\Vert \Xi( \rho_{1} -\rho_{2} ) \Vert_{1} \over \Vert \rho_{1} - \rho_{2}\Vert_{1}}  \nonumber \\ &=& {\Vert \Xi( \rho_{1} ) -e^{i\pi a^{\dagger}a}\Xi(\rho_{1} )e^{-i\pi a^{\dagger}a} \Vert_{1} \over \Vert \rho_{1} - \rho_{2}\Vert_{1}}. \label{step1}
\end{eqnarray}

From the characterization of $\Vert \rho - \sigma \Vert_{1}$ as the maximal $\ell_{1}$ distance between the images of $\rho$ and $\sigma$ under a quantum measurement, we find that, for a given positive operator-valued measure $\lbrace E(dx) \rbrace$ defining the quantum measurement with outcome space $\mathcal{X}$,
\begin{eqnarray}
\Vert \rho_{1} - \rho_{2}\Vert_{1} \tau(\Xi) &\ge & \int \Big\vert \mathrm{tr}\left(\vphantom{e^{-i\pi a^{\dagger}a}} E(dx)\Xi( \rho_{1} ) \vphantom{e^{-i\pi a^{\dagger}a}}\right) \nonumber \\ &-& \mathrm{tr}\left( E(dx) e^{i\pi a^{\dagger}a}\Xi(\rho_{1} )e^{-i\pi a^{\dagger}a} \right) \Big\vert . \nonumber \\ &{}&  \label{step2}
\end{eqnarray}
By taking $\mathcal{X}=\mathbb{C}$ and choosing positive operator-valued measurement $\lbrace \ket{\alpha} \bra{\alpha} {d^{2}\alpha \over \pi} \rbrace$, Eq.(\ref{step2}) becomes

\begin{equation}
\tau(\Xi) \ge { \int {d^{2}\alpha \over \pi} \big\vert Q_{\Xi(\rho_{1})}(\alpha) - Q_{\Xi(\rho_{1})}(-\alpha)\big\vert \over \Vert \rho_{1} - \rho_{2}\Vert_{1} } , \label{step3}
\end{equation}
where, for a quantum state $\rho$, $Q_{\rho}(\alpha):= \langle \alpha \vert \rho \vert \alpha \rangle$ is the Husimi Q function. Note that a corollary of the construction of $\ket{\Omega_{+}}$ in Section \ref{sec:supdef} is that each diameter of $CCH(G(\mathcal{H})_{E})$ is defined by a pair of pure Gaussian states that are identical, except for having opposite mean vectors. Therefore, the right hand side of Eq.(\ref{step3}) can be used to compute a lower bound for the trace norm contraction coefficient for every diameter of $CCH(G(\mathcal{H})_{E})$. More generally, the bound in Eq.(\ref{step3}) holds for any quantum channel covariant with respect to a group that contains a $\mathbb{Z}_{2}$ subgroup.


\section{Conclusion}
By identifying the pairs of pure Gaussian states $\ket{\varphi_{1}}$, $\ket{\varphi_{2}}$ that exhibit maximal trace distance subject to an energy constraint, and proposing to utilize the lowest energy state $\ket{\Omega_{+}}$ in the span of $\ket{\varphi_{1}}$, $\ket{\varphi_{2}}$ as the initial environment state in the Stinespring form of a quantum channel, we have motivated a distinct class of linear bosonic quantum channels. Unlike the case of linear bosonic Gaussian channels, the present channels do not map the set of Gaussian states to itself. 

We examined the transfer of nonclassical features to the Fock vacuum by the $\Xi_{\zeta = \pi /4}$ channel and compared the nonclassicality distance of $\Xi_{\zeta = \pi /4}(\ket{0}\bra{0})$ to the value obtained when the environment is initialized in an even coherent state. In particular, we found that the $\Xi_{\zeta=\pi/4}$ channel maps a single mode system prepared in a coherent state to a nonclassical state even for asymptotically large values of the environment energy constraint. Furthermore, we computed the critical thermal noise beyond which the $\Xi_{\zeta = \pi/4}$ channel cannot convert a classical system to a nonclassical state. Finally, the second moment noise generated by the $\Xi_{r}$ phase insensitive linear amplifier was calculated both for systems initialized in a coherent state and in $\ket{\Omega_{+}}$, and a general lower bound for the trace norm contraction coefficient of the $\Xi$ channels was derived.

In this paper, no attempt has been made toward a detailed mathematical description (e.g., derivation of a normal form) of the structure of the set of linear bosonic quantum channels defined by superpositions of maximally distant, isoenergetic environment states. Future work could involve an analysis of the communication capacities of the $\Xi$ channel, which are closely related to the entropy production by the channel and its complement. Progress in this direction has already been made in the case of photon-added linear bosonic channels, which are defined by linear quantum dynamics and an environment initialized in a non-vacuum Fock state \cite{wintersaba}. Investigations into these topics would constitute further progress toward understanding the general properties non-Gaussian linear bosonic channels. We expect the present work to provide a basis for such studies.

\nonumsection{Acknowledgements}
\noindent
The author thanks Yongkyung Kwon for hosting during the completion of this work. This work was supported by the Korea Research Fellowship Program
through the National Research Foundation of Korea (NRF) funded by the Ministry of Science and ICT
(2016H1D3A1908876) and by the Basic Science Research Program through the National Research
Foundation of Korea (NRF) funded by the Ministry of Education (2015R1D1A1A09056745).

\nonumsection{References}
\noindent

\bibliographystyle{unsrt}
\bibliography{chanbib}

\appendix

\section{Selected properties of $\ket{\Omega_{+}}$ \label{sec:app1}}

Several properties of superpositions of ``ideal coherent states''  of the form $\ket{(\alpha , r)} + \ket{(-\alpha ,r)}$, with $\alpha \in \mathbb{C}$ and $r\in \mathbb{R}$ were calculated in Ref.\cite{moyacessasqueezing}. Superpositions of tensor products of such states have been considered as generalizations of the well-known entangled coherent states and hierarchical Schr\"{o}dinger cat states \cite{volkoffcat}. Here, we collect some facts concerning $\ket{\Omega_{+}}$ and the channel $\Xi$ that are relevant to the applications in the main text.

\subsection{Fock basis amplitudes\label{sec:fock}}

\begin{figure} [htbp]
\centerline{\epsfig{file=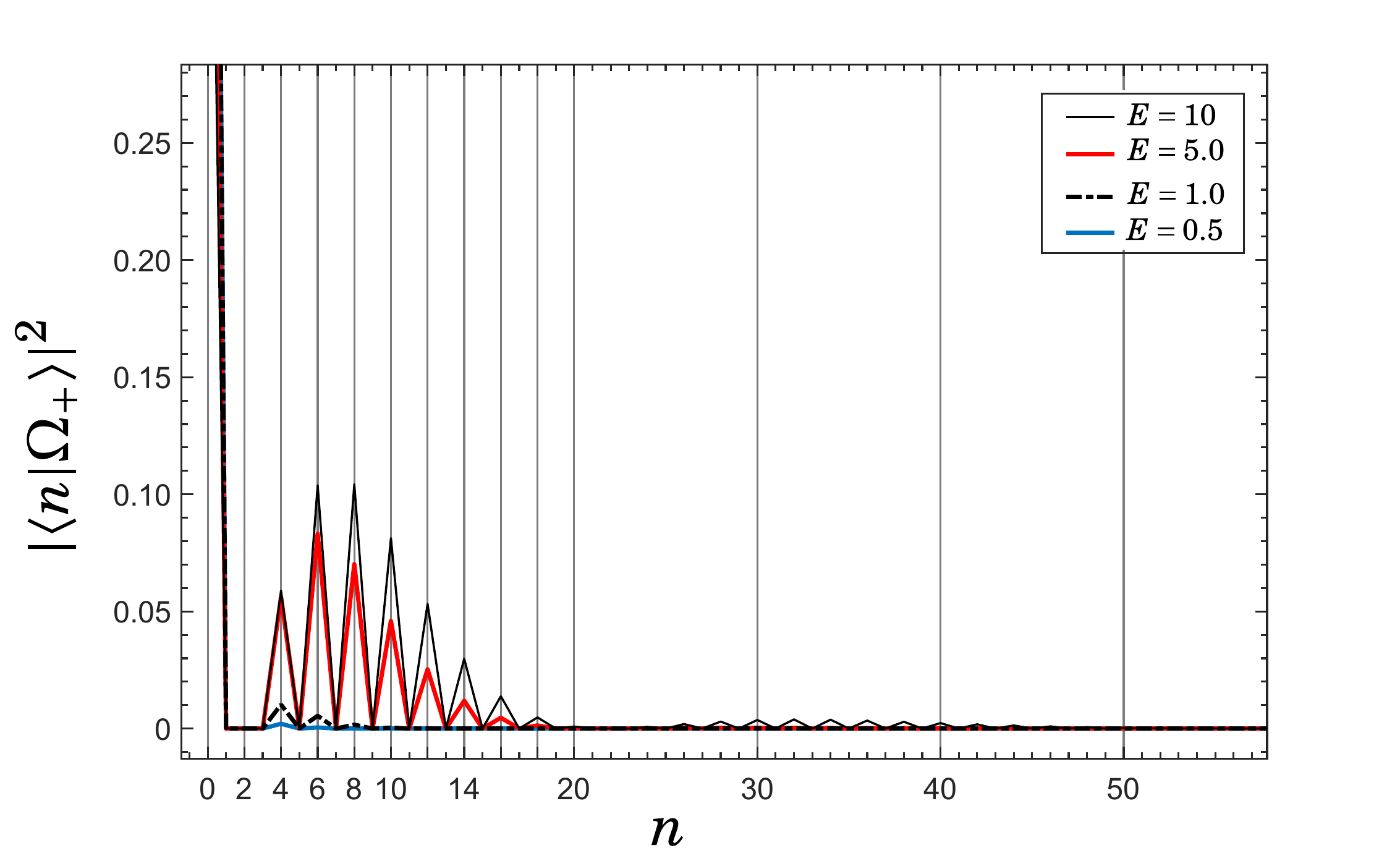, width=8.2cm}} 
\vspace*{13pt}
\fcaption{\label{fig:omegaprops}a) Fock state distribution for $\ket{\Omega_{+}}$ for $E=0.5,1.0,5.0,10$. The vacuum probability $p(0):=\vert \langle 0 \vert \Omega_{+} \rangle \vert^{2}$ is not shown; the corresponding values are: $(E=0.5, p(0)= 0.9974)$,  $(E=1.0, p(0)= 0.9822)$, $(E=5.0, p(0)= 0.6987)$, $(E=10, p(0)= 0.5175)$. }
\end{figure}

The amplitudes $\langle n \vert \Omega_{+} \rangle $ of $\ket{\Omega_{+}}$ in the Fock state basis $\lbrace \ket{n} : n=1,2,\ldots \rbrace$ can be obtained straightforwardly from the analogous amplitudes for the two-photon coherent state (Ref.\cite{mandel}, p. 1050). The Fock state basis amplitudes for the state $\ket{\Psi}:=S(w)\left( {D(\gamma) + D(-\gamma) \over \sqrt{2 + 2\exp{(-2\gamma^{2})}} } \right)\ket{0}$, with $\gamma$, $w \in \mathbb{R}$ are:

\begin{equation}
\langle n \vert \Psi \rangle =
\left\{
        \begin{array}{ll}
            0 & \quad ,\,n \; \mathrm{odd} \\
            { \left( {\tanh w \over 2} \right)^{n/2}e^{{1\over 2}\gamma^{2}\tanh w} \over \sqrt{n!}\sqrt{\cosh(w) \cosh (\gamma^{2}) }}H_{n}\left( {\gamma \over \sqrt{\sinh 2w} } \right) & \quad ,\, n \; \mathrm{even}
        \end{array}
    \right.
    \label{famp}
\end{equation}

where $H_{n}(x)$ are the Hermite polynomials. Substituting $w={1\over 2}\ln d_{c}(E)$, $\gamma = \gamma_{c}=r(d_{c}(E))\sqrt{d_{c}(E)}$ into Eq.(\ref{famp}) gives the $E$-dependent Fock state amplitudes for $\ket{\Omega_{+}}$. The Fock state distribution $\vert \langle n \vert \Omega_{+} \rangle \vert^{2}$ is shown for various values of $E$ in Fig. \ref{fig:omegaprops}. The number distribution corresponding to the optimal parameters of $\ket{\Omega_{+}}$ exhibits a notable feature, namely that $\langle n \vert \Omega_{+} \rangle = 0$ for $n=2$. This fact is obtained from Eq.(\ref{famp}) by substitution of $x=2\gamma_{c} /\left( d_{c}(E) + (d_{c}(E)^{-1})\right) $ into $H_{2}(x)=4x^{2}-2$. The Fock state ``revivals'' exhibited for $E\gtrsim 5$ constitutes a second notable feature of the number distribution.

\subsection{\label{sec:altexpr}Alternative expressions for $\ket{\Omega_{+}}$}
For ease of calculation, it is useful to note that the state $\ket{\Omega_{+}}$ can be expressed as a superposition of distributions of coherent states on the lines $C_{\pm}$ in the complex plane defined by the equations $z=\pm r(d_{c}(E))+ix$, respectively, where $x\in (-\infty , \infty)$. This property follows from the analogous property of the squeezed vacuum and the even coherent state \cite{janszky,janszkycomment,janszkyreply}. For the normalized state $\ket{\psi(r,d)} \propto \ket{ (r, {1\over 2}\ln d )} + \ket{ (-r, {1\over 2}\ln d )} $, where $r\ge 0$ and $d >1$, the integral expression for the normalized state is found to be:
\begin{equation}
\ket{\psi(r,d)} = {d^{1\over 4} \over \sqrt{ \pi (d-1)\left( 2+2e^{-2dr^{2}} \right) }} \int_{-\infty}^{\infty}dx \, e^{-\left( 1 \over d-1 \right)x^{2}} \left[ e^{-irx}   \ket{ix+r}+  e^{irx}   \ket{ix-r} \right] .
\label{linesuper}
\end{equation}

We define the characteristic function $\chi_{\rho}: \mathbb{R}^{2} \rightarrow \mathbb{C}$ of a state $\rho$ by $\chi_{\rho}(x,y) = \mathrm{tr}\left( \rho e^{ixq + ipy} \right)$, where $q$ and $p$ are the position and momentum operators satisfying $[q,p]=i$. For a state of the form $\ket{\Psi}$ defined in Section \ref{sec:fock}, e.g., $\ket{\Omega_{+}}$, the characteristic function is given by
\begin{equation}
\chi_{\ket{\Psi}\bra{\Psi}}(x,y) = {e^{-{1\over 4}\left( e^{-2w}x^{2} + e^{2w}y^{2} \right) } \over 1+e^{-2\gamma^{2}}}   \left[ \cos \left( \sqrt{2} \gamma e^{-w} x \right) + e^{-2\gamma^{2}}\cosh \left( \sqrt{2} \gamma e^{w} y \right) \right].
\label{optcharfunct}
\end{equation}

\section{Operator-sum representation of $\Xi$ \label{sec:app2}}

To relate the expression of the quantum channel $\Xi$ given in Eq.(\ref{genchan}) to the formalism of quantum operations \cite{davies}, it is necessary to derive an operator-sum representation for $\Xi$. A discrete operator-sum representation of the channel $\Xi$ can be derived from the  Fock basis  matrix elements $\langle n',m' \vert U_{\mathrm{BS}/\mathrm{TM}} \vert n,m \rangle$ of $U_{\mathrm{BS}/\mathrm{TM}}$ and the Fock basis amplitudes $\langle n \vert \Omega_{+} \rangle$ of the environment state using the technique of Ref.\cite{sabapathyopsum}. To derive a continuous operator sum representation of the form $\Xi(\rho)=\int {d^{2}\beta \over \pi} \, K_{\beta}\rho K_{\beta}^{\dagger}$, one must simplify the operator $K_{\beta}:= {}_{E}\langle \beta \vert U \vert \Omega_{+} \rangle_{E}$, where $U=U_{\zeta}$ or $U=U_{r}$, which can be achieved by the appropriate Lie group factorizations (SU(2) for $U_{\zeta}$ and $SU(1,1)$ for $U_{r}$) and use of the coherent state representation of $\ket{\Omega_{+}}$ (Eq.(\ref{linesuper})).

\end{document}